# IDENTIFICATION OF MOST SIGNIFICANT PARAMETER FOR OPTIMIZING THE PERFORMANCE OF RPL ROUTING PROTOCOL IN IoT USING TAGUCHI DESIGN OF EXPERIMENTS

By

**CHANDRA SEKHAR SANABOINA ***      **PALLAMSETTY SANABOINA ***

*\* Research Scholar, Jawaharlal Nehru Technological University Kakinada, Andhra Pradesh, India.*
*\*\* Professor, Department of Computer Science and Systems Engineering, Andhra University, Visakhapatnam, Andhra Pradesh, India.*

## ABSTRACT

*Internet of Things (IoT) consists of a wide variety of devices with limited power sources. Due to the adhered reason, energy consumption is considered as one of the major challenges in the IoT environment. In this research article, an attempt is made to optimize the existing Routing Protocol (RPL) towards a green technology. It focuses on finding the most significant parameter in the RPL using Taguchi Design of Experiments. It emphasizes the effects of five input factors, such as Network Size, Mobility Speed, DIO_DOUBLING, DIO_MIN_INTERVAL, and Redundancy Constant on only one output parameter Power Consumption. The findings show that DIO_MIN_INTERVAL is the leading factor that has a significant effect on the power consumption in RPL. After determining the most significant factor that affects the power consumption, measures can be taken to optimize the performance of RPL by applying some optimization techniques. COOJA simulator is used to carry out the simulations required for this research article.*

## INTRODUCTION

A Wireless Sensor Network (WSN) consists of small devices which are in general called as nodes. These nodes are embedded with sensor devices, actuator devices, a computational unit for processing the sensor data, and optionally communication unit for communicating the processed data to remote locations. The WSNs finds its applications in environmental monitoring. Researchers found their interest in WSNs with the advent and rapid advancement in technological aspects of wireless technology and embedded systems.

Various components in WSN nodes perform crucial tasks by making nodes able to communicate with each other to transmit data obtained by their sensors. Communication may take place internally (i.e., between nodes) or externally (i.e., between the node and a centralized system). The various communication modes in WSNs had lead to the development of the conception of the Internet of Things (IoT). With the advent of IoT, one can be able to access efficient and accurate environmental data immediately. This will increase productivity because the user is able to access and analyze the most up-to-date information.

WSN can be defined as a network of nodes that work in a cooperative way to sense and control the environment surrounding them. These nodes are linked via wireless media. Nodes use this connection to communicate with each other. Conceptually, a WSN is based on a simple equation which depends on the fact that Sensing + CPU + Radio = Lots of Potential (Ephrem, 2015). Sensing Unit is necessary to monitor the surrounding environment and its conditions, such as humidity, pressure, and vibration. After completing monitoring and sensing processes, necessary computations are accomplished in the CPU. Lastly, computed environmental data are transferred from Radio

Unit through the wireless communication channels among the nodes. Finally, these data are sent to the Gateway.

The inception of WSNs research took place in the 1980s, but from 2001 onwards the use of WSNs was extensively found in the industries. The reason for this might be due to the advent of miniature technologies like System on Chip (SoC) and Very Large Scale Integrated Circuits (VLSI), which reduced the size of sensors, processors, and other components to an unbelievably small size.

IEEE organization has defined a standard for this fact. The IEEE 802.15.4 covers low data rate wireless personal area networks. Based on this standard, ZigBee Alliance has published the ZigBee standard that can be used in WSNs.

The idea of the Internet of Things (IoT) was developed in parallel to WSNs. The term internet of things was coined by Ashton (2009). It refers to the process of giving a unique address to each and every object (Things) available in the real world thus making them available over the internet. These things can vary from large entities like airplanes, Industrial plants, machines, cars to small entities like spectacles, pens, etc., or even they can be the body parts of human beings, animals, and plants. Wireless communication technologies play a major role in communication between things. Even though not strictly confined to this wireless technology it can also extend to wired communications.

According to Cisco (Evans, 2011), 50 billion things will be connected to the Internet in 2020, thus overshadowing the data generated by humans. This is limited by the birth rate in 2020. In this study, for optimizing the existing Routing Protocol (RPL) towards a green technology and finding the most significant parameter in RPL, Taguchi approach was used.

*Taguchi Design of Experiments (DoE)*

Design of Experiments (DoE) is a multi-purpose technique (Box et al., 2005). Its usage is not limited to physical experiments, but can be applied to simulation experiments (Law & Kelton, 2000), to the investigation of calculated results of complex analytical expressions whose parameters are methodically varied or to other decision problems, where the effects of several factors are examined. DoE is also known as the factorial design of experiments was first proposed by R. A. Fisher. As the name suggests, the full factorial design of experiments identifies all the possible combinations for the given set of factors and hence results in a very large set of experiments. In fact, all the combinations of the experiments need to be conducted. As an example, consider industrial experiments where almost every factor is considered to be important. In such cases, the full factorial design of experiments results in a large number of experiments which incur a lot of experimental costs. To put a check on the experimental costs, proposals were made to reduce the number of experiments to a practical level but still, give efficient and accurate results. This new proposal with a limited number of experiments is branded as the partial fraction design of experiments. The drawback with this proposal is that it does not have general guidelines for its application. It is an informal way of conducting experiments and no one can guarantee and authenticate the results obtained from this method.

The next proposal for conducting the experiments is based on a special set of general design guidelines which was proposed by Taguchi. These guidelines help the researchers to conduct experiments covering a wide area of applications. Taguchi DoE was developed in the 1960s in Japan and was scattered to the west during the 1980s. This Taguchi style has its own advantages and disadvantages (Taguchi, 1986; Dehnad, 1989). Orthogonal arrays are a special set of arrays used in this method, which specifies the conduction of a minimal number of experiments, but can give full fledged information of all the factors that affect the parameter under consideration. The DoE using the orthogonal array is, in most cases, efficient when compared to many other statistical designs. The minimum number of experiments that are required to conduct the Taguchi method can be calculated based on the degrees of freedom approach. This article mainly focuses on finding the most significant parameter in RPL for the optimization of power consumption.

1. Routing Protocol for Low-Power and Lossy Networks (RPL)

The ROLL (Routing Over Low Power and Lossy Network) is a working group created by Internet Engineering Task Force (IETF) for the purpose of analyzing the routing requirements of applications including industrial automation, home and building automation, and smart grid. The main objective of ROLL working group was to design and develop the routing solutions for IP-based Low power and Lossy Networks (LLN) that have the support of a variety of link layers. An LLN is made up of embedded devices that have limited memory, low energy (battery power), and low processing capability.

RPL is a proactive routing protocol based on IPv6 distance vector protocol. Construction of DODAG is the first step in the routing process and routing the data packets through the DODAG is the second step. The name DODAG suggests its inherent property that all edges are oriented towards the destination and no cycles exist. Every node in the DODAG is designated with a rank, which specifies the relative position of that particular node with respect to the DODAG root node. It is obvious that the rank of the node strictly increases from the root towards the leaf nodes.

The rank is computed depending on the DODAG's Objective Function (OF): hop counts, link metrics (ETX, i.e. the expected number of transmissions required to successfully transmit and acknowledge a packet on the link or LQI, i.e. the Link Quality Indicator) or other constraints. To build and maintain its logical topology (route, parents, neighbors table), RPL uses IPv6 control messages (Tsvetkov & Klein, 2011).

### 1.1 Trickle Algorithm

Power consumption is a major trade in resource-constrained environments. As per the research findings, the major source of energy consumption in RPL is due to the transmission of the control packets. Careful transmission of the control packets will lead to green technology (i.e., less power consumption). There exists an algorithm that controls/limits the transmission of a number of control packets. The time interval between two consecutive Data Input Output (DIO) messages is considered to be the DIO minimum interval. This DIO minimum interval doubles after each successful transmission of the DIO message and is limited by a maximum value, which is determined by DIO interval doublings. Three parameters that are configurable and has a major influence on the performance of the RPL are $I_{min}$, $I_{max}$ and Redundancy Constant (K) (Winter et al., 2012).

### 1.1.1 $I_{min}$ Parameter

It is the minimum interval of time between two successful DIOs. In general, DIO control messages are periodically transmitted so that optimal use of resources can be achieved. The periodical transmission also reduces the redundant control traffic. An algorithm keeps control of the transmission of the DIO messages which is named as a Trickle Timer algorithm. It has minimum and maximum values, which are dictated by $I_{min}$ and $I_{max}$ parameters, respectively. The trickle timer starts at the lowest possible value $I_{min}$ and gets doubled each time a DIO message is successfully transmitted until it reaches a maximum possible value of $I_{max}$.

Calculation of $I_{min}$ is given by the formula.

$I_{min} = 2 \; \hat{} \; RPL\_DIO\_INTERVAL\_MIN$

where RPL_DIO_INTERVAL_MIN is an RPL parameter in Contiki Operating System.

### 1.1.2 $I_{max}$ Parameter

This parameter is used to limit the number of times the $I_{min}$ can be doubled. The value of $I_{max}$ is determined by the RPL parameter DIO Interval Doublings (In Contiki: RPL_DIO_INTERVAL_DOUBLINGS) and computed as:

$I_{max} = I_{min} * 2 \; \hat{} \; RPL\_DIO\_INTERVAL\_DOUBLINGS$

### 1.1.3 Redundancy Constant

The average transmission probability of a node in the network depends on the number of neighbors and the redundancy constant. The usage of a fixed redundancy constant in the network causes an unbalanced transmission load and may cause early depletion of energy sources of nodes with fewer neighbors. Increasing the redundancy constant increases the transmission probabilities in the network (Coladon et al., 2015).

Redundancy constant (k) is a natural number greater than 0 and is used to suppress the DIO transmission (Hazrat, 2012).

### 1.1.4 DIO Interval Minimum

This parameter controls the rate of DIO transmission and therefore crucial for control traffic overhead, energy consumption, convergence time, etc. The more quickly the DIOs are transmitted the more quickly the network gets converted, but at the expense of Control Traffic overhead and more energy consumption.

This parameter is influential on the performance of the whole protocol performance. A careful tweaking of this parameter is necessary for improved performance keeping in view the different application areas of WSN and environmental conditions.

*1.1.5 DIO Doublings*

Maintaining the steady network conditions and low traffic is a crucial part in RPL. Hence, the DIO min interval is doubled after each successful DIO transmission and the parameter is defined as DIO doubling.

Its low value can cause the control traffic to flow even if not needed. Consequently, it is essential to configure this parameter precisely.

*1.1.6 Network Size*

The number of nodes in a network defines the network density. Network size (N/W) is one of the factors that affect the performance of the RPL routing protocol. In this study, different network densities are evaluated to discover the importance of nodes and how they can influence RPL behavior (Kumar et al., 2011).

*1.1.7 Mobility Speed*

There are many aspects of routing in IoT which are harder to deal with when nodes are mobile because the issues like energy efficiency, PDR, and connectivity become more difficult to optimize. Solutions for mobility usually rely on updating routing information frequently, which is a critical factor for IoT with constrained resources. Therefore the evaluation of RPL with mobile nodes with different specifications provide good suggestions and improvements for further enhancements of the routing protocol (Hazrat, 2012).

## 2. Methodology

The methodology used in this research article for performance evaluation of RPL is explained in this section.

*2.1 Concepts of Taguchi DoE*

This is a new method of conducting experiments which are based on well-defined guidelines. Taguchi's effort of designing a set of guidelines for conducting the experiments was named after him and is popularly known as Taguchi Design of Experiments. Two vital tools are utilized in Taguchi design: Orthogonal Arrays (OAs) and Signal-to-Noise Ratio (SNR). OAs are a special type of arrays which give a combination of the factors with a limited number of experiments, but could confer full information that affects the performance parameter (Mohamed et al., 2010).

The bottom line of OAs lies in choosing various levels of combinations of the input factors for each experiment. Degrees of freedom approach is used to find the optimal minimum number of experiments that need to conduct the Taguchi DoE. For any OA, the degrees of freedom should be greater than or equal to the number of factors under consideration (i.e., input parameters) (Mohamed et al., 2008).

$$N_{Taguchi} = 1 + \sum_{i=1}^{NV} (Li - 1) \qquad (1)$$

The variation which affects the change in performance of RPL is computed through Signal-to-Noise Ratio (SNR). The SNR is used to measure the performance metrics as well as the significant parameters through Analysis of Variance (ANOVA). Three classes of the performance metric in the analysis of SNR are employed: larger-the-better, smaller-the-better, and the nominal-the-best.

*2.1.1 Smaller-The-Better*

In this research article, the smaller-the-better metric is chosen for performance evaluation of the chosen parameters of RPL. This metric is used to analyze the power consumptions (Roy, 2001). The SNR is computed as:

$$N_p = 10 \log(1/r \sum_{i=1}^{r} y^2) \qquad (2)$$

where r is the number of simulation repetitions under the same design point, p is the power, and y is the response value.

*2.1.2 Larger-The-Better*

Larger-the-better metric is used for analyzing the strength, efficiency, S/N ratio, and throughput of the routing protocols.

$$N_p = -10 \log(1/r \sum_{i=1}^{r} \frac{1}{y^2}) \qquad (3)$$

*2.1.3 Nominal-The-Best*

In this metric, the signal value is fixed (nominal value), and the variance around this value can be considered the result of Signal-to-Noise factors.

$$N_p = 10 \log_{10}(\alpha^2) \qquad (4)$$

where $\alpha$ is the mean value.

*2.1.4 ANOVA*

The Analysis of Variance (ANOVA) is used to determine the relative effect of all factors on the signal metric (i.e, power consumption) and to determine which factor has the highest effect.

Parameters in ANOVA are calculated by the following equations. The Sum of Squares ($SS_T$) from SNR (Voice et al., 2007; Sahithi & Setty, 2018) is given as follows:

$$SS_T = \sum_{I=1}^{N}(N_i - \bar{N})^2 \qquad (5)$$

where $N_i$ is the Signal to Noise Ratio (SNR) of the $i^{th}$ experiment conducted.

The sum of squared deviations due to each factor ($SS_j$) is calculated by using the following formula:

$$SS_I = \sum_{I=1}^{L}(N_{ji} - \bar{N})^2 \qquad (6)$$

where L is the number of levels and $N_{ji}$ is the average SNR of a $j^{th}$ factor at $j^{th}$ level. Also, the sum of squares of error (SSc) is given by,

$$SS_C = SS_T - \sum_{j=1}^{q} SS_j \qquad (7)$$

where q represents the number of factors, $SS_j$ represents the sum of squared deviations for each factor. The contribution of each factor $p_j$ in percentage is given by the formula:

$$p_j = (SS_J / SS_T) * 100 \qquad (8)$$

Where $p_j$ value gives the significance level for each factor. The f-test can also be used to determine which factor has the most significant effect on the performance metric. The large F-ratio indicates the strong effect of the factor. The F-value for $j^{th}$ can be evaluated as:

$$F = (SS_j / Df_j) / (SS_c / Df_c) \qquad (9)$$

*2.2 Designing Taguchi Technique*

It becomes impossible to utilize the conventional design of experiments because of its highly unpredictable behavior. The process parameters are directly proportional to the number of experiments (i.e., the number of experiments to be conducted increases with the increase in the number of input factors/parameters). The Taguchi method finds a solution to this problem. It makes use of a unique type of arrays called orthogonal arrays which studies the entire parameter space with a limited number of experiments. The steps required for designing a Taguchi experiment is illustrated in Figure 1.

*2.2.1 Selection of the Independent Factors*

Identifying the factors that most influence the output parameter has significant importance. This knowledge is essential prior to conducting the experiment. The input to the experiment is generally obtained from the project which benefits in compiling a widespread list of factors.

*2.2.3 Deciding the Number of Levels*

After identifying the independent factors, the user has to decide upon the number of levels (number of values) for each factor. The choice of different levels is dependent on the effect of the output parameter due to different levels of the input parameters.

*2.2.3 Selection of Orthogonal Array*

The degrees of freedom plays an important role in deciding the orthogonal array. A minimum number of experiments that are to be conducted depends on the selection of orthogonal array. It is important to note that the degrees of freedom available should be less than the minimum number of experiments that must be run.

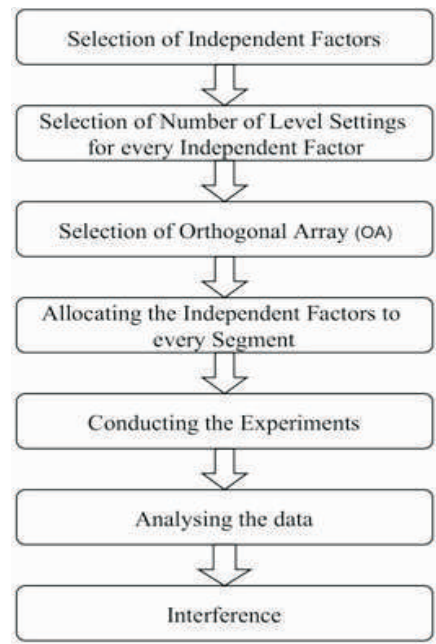

Figure 1. Taguchi's Methodology

*2.2.4 Allocating the Independent Factors to every Segment*

Independent factors are to be assigned to the vertical column and the order of assignment is very crucial. The factors are to be assigned at right columns in case of mixed level factors (i.e., different factors having different levels). Deciding upon the actual level of values for each design factor is essential before conducting the experiment. The percentage contribution and the importance of the independent factors change based on the number of values (levels) assigned to them. Setting the proper values to various independent factors is the designer's liability.

*2.2.5 Conducting the Experiments*

Various experiments suggested by the orthogonal array as per the level combinations are to be performed. It is mandatory that all the experiments suggested by the OA should be conducted. The dummy factor and interaction columns are needed only at the time of analyzing the data and are to be omitted during the conduction of experiments. At the end of each experiment, the parameter under consideration is to be noted down.

*2.2.6 Analysis of the Data*

Segregation of individual effect of independent factors is essential as each experiment is the combination of various factors with different levels. Segregation is the process of summing up the values of the parameter under consideration for different levels. After finding the mean value of each level of an independent factor, the sum of the square of deviation of each of the mean value from the total mean value is to be calculated.

Whether the parameter under considerations is sensitive to the change in level setting is decided by the sum of square deviation. The zero value or close to zero value of the sum of square deviation indicates that the design factors are insignificant (i.e., factors are not influencing the performance of the process). Analysis of Variance (ANOVA) can be conducted to decide upon which independent factor has highest percentage contribution.

*2.2.7 Interference*

It is clear from the above experimental analysis that smaller the value of the sum of the square of an independent factor the smaller will be its influence of the parameter of consideration and more the value of the sum of the square of an independent factor the more will be its influence on the parameter under consideration. It is good to calculate the ratio of the individual sum of the square of a factor to the total sum of squares of all the factors. This ratio emphasizes the percentage contribution of the independent factors on the parameter under consideration.

3. Simulation Environment

In this section, the simulation environment which was used for the performance evaluation of RPL is explained.

*3.1 Contiki Operating System*

The Operating System that is best suited for the wireless sensor network operating system is Contiki. It has the inbuilt kernel, libraries, the program loader, and various set of processes in the kernel as well as for the user (Dunkels et al., 2004). This operating system is most widely used in the networked embedded systems and smart environment, which involves a large number of smart objects. It provides a lot of means that assist in writing the applications for smart objects. Libraries provide a means which assists in linked list manipulation, memory allocation, and communication with the external world. Perhaps, Contiki is the first embedded Operating System that provided IP communication. 'C' is the programming language that is used to develop the Contiki operating system. All the applications in Contiki were also developed using C programming language and hence it supports portability to different architectures like Texas Instruments MSP430.

Processes in Contiki are implemented as event handlers and hence Contiki can be treated as an event-driven operating system. The main components of the Contiki operating systems are core and loaded programs.

- *Core:* It consists of Contiki Kernel, the language run-

time, the program loader, and a communication stack with device drivers for the communication.

- *Loaded Programs:* It consists of binary image files of sensor motes which are used in Cooja simulator (Dunkels et al., 2004).

### 3.2 Cooja Simulator

An inbuilt simulator that is available in Contiki sensor network operating system is Cooja. It is based on Java (Osterlind et al., 2006). The simulator is entirely developed using Java, but some sensor node software is written in 'C' programming language. One of the outstanding features of Cooja is that it allows simultaneous simulations at Network Level, Operating System Level, and Machine Code Instruction Level (Osterlind et al., 2006). Cooja has the capability of running Contiki programs compiled for native CPU or MSP430 emulator.

Plugins like Simulation, Visualizer, Radio Logger, and Timeline are used to implement interactions with the simulated nodes. All the simulation data is stored in an XML file, which has an extension of CSC (Cooja Simulation Configuration). It consists of all the information related to nodes, node position, random seed, radio medium used, Simulation environment, etc.,

The Cooja simulator has the full control of the simulated systems because of the approach if followed during simulation. It makes use of functions for analyses and control of the Contiki system. As an example, consider a situation where the simulator informs the Contiki system to handle an event or fetches the entire Contiki system memory for analysis.

The drawbacks of the Cooja simulator is the use of JNI. It leads to some of the side effects. Another important drawback is the dependency on external tools like compilers, linkers, and their run-time arguments.

### 3.3 Experimental Design and Simulation

In this article, the performance of RPL is evaluated based on five factors: Network Size, Mobility Speed, DIO_MIN_INTERVAL, DIO_DOUBLING, and Redundancy Constant as described in Table 1. As listed, each parameter is examined at three different levels (i) Level-1, (ii) Level-2, and (iii) Level-3. Based on the number of parameters and number of levels, $L_{27}$ equals to $3^5$ ($Level^{factors}$) = 243 experiments have to be conducted if it is a Full-Factorial method.

| Label | Factor | Level 1 | Level 2 | Level 3 |
|-------|--------|---------|---------|---------|
| A | Network Size (NS) | 20 Nodes | 30 Nodes | 40 Nodes |
| B | Mobility Speed (MS) | 5 m/s | 15 m/s | 25 m/s |
| C | DIO_MIN_INTERVAL (MIN) | 8 | 12 | 16 |
| D | DIO_DOUBLING (DOUBLING) | 4 | 8 | 12 |
| E | REDUNDANCY_ CONSTANT (RC) | 6 | 10 | 14 |

Table 1. Experimental Parameters and their Levels

In a fixed level (3 level design), two orthogonal arrays $L_9$ and $L_{27}$ are available. Selecting $L_9$ needs only 9 experiments to be conducted which may not give sufficient results for correct analysis. Therefore, $L_{27}$ OA is selected in this research article and 27 experiments were conducted using Cooja simulator with different levels of the factors.

Table 2 depicts the combination of different experiments and each combination of the experiment can be termed as Design-Point and each Design-Point corresponds to a simulation scenario.

Table 3 gives an idea of the simulation environment used to conduct experiments. This research emphasizes a small simulation area of size 100 m × 100 m. Simulations (Design-Point) are carried out for a period of 600000 ms (600 s). The experiments were repeated thrice to check whether the simulation parameters selected (i.e, simulation area 100 m × 100 m and simulation time 600000 ms) were optimum or not. The three set of results for various design points generated the same results thus concluding that the selected simulation parameters are apt and hence steady state has been achieved.

As each simulation scenario is executed, the CPU Power, LPM Power, Listen Power, Transmit Power, and Overall Power is computed as output performance metrics. But for evaluation purpose of this research article, only overall average power is considered because overall average power is treated as:

Overall Average Power ≅ CPU + LPM + Radio Listen + Radio Transmit Power

### 4. Results and Data Analysis

Figure 2 depicts the main effects plot of S/N ratio for power.

| Design Point | Level of Factors | | | | |
|---|---|---|---|---|---|
| | A | B | C | D | E |
| 1 | 20 | 5 | 8 | 4 | 6 |
| 2 | 20 | 5 | 8 | 4 | 10 |
| 3 | 20 | 5 | 8 | 4 | 14 |
| 4 | 20 | 15 | 12 | 8 | 6 |
| 5 | 20 | 15 | 12 | 8 | 10 |
| 6 | 20 | 15 | 12 | 8 | 14 |
| 7 | 20 | 25 | 16 | 12 | 6 |
| 8 | 20 | 25 | 16 | 12 | 10 |
| 9 | 20 | 25 | 16 | 12 | 14 |
| 10 | 30 | 5 | 12 | 12 | 6 |
| 11 | 30 | 5 | 12 | 12 | 10 |
| 12 | 30 | 5 | 12 | 12 | 14 |
| 13 | 30 | 15 | 16 | 4 | 6 |
| 14 | 30 | 15 | 16 | 4 | 10 |
| 15 | 30 | 15 | 16 | 4 | 14 |
| 16 | 30 | 25 | 8 | 8 | 6 |
| 17 | 30 | 25 | 8 | 8 | 10 |
| 18 | 30 | 25 | 8 | 8 | 14 |
| 19 | 40 | 5 | 16 | 8 | 6 |
| 20 | 40 | 5 | 16 | 8 | 10 |
| 21 | 40 | 5 | 16 | 8 | 14 |
| 22 | 40 | 15 | 8 | 12 | 6 |
| 23 | 40 | 15 | 8 | 12 | 10 |
| 24 | 40 | 15 | 8 | 12 | 14 |
| 25 | 40 | 25 | 12 | 4 | 6 |
| 26 | 40 | 25 | 12 | 4 | 10 |
| 27 | 40 | 25 | 12 | 4 | 14 |

Table 2. Experimental Layout using $L_{27}$ Orthogonal Array

| Routing Protocol | RPL |
|---|---|
| Simulation Time | 600000 ms |
| Simulation Area | 100 m × 100 m |
| Network Size | 20, 30, 40 Nodes |
| Mobility Speed | 5, 15, 25 m/s |
| MIN | 8, 12, 16 |
| DOUBLING | 4, 8, 12 |
| RC | 6, 10, 14 |

Table 3. Simulation Parameters

In the plots given, Y-axis indicates the dependent variables (Mean SNR in this case) and X-axis indicates Independent variables (NS, MS, MIN, DOUBLING, and RC). The main effects plot gives the effect of a single independent variable on a dependent variable ignoring all other independent variables. The main effects plot is an easy decision tool that helps in understanding the data means. The analysis of the main effects plot for power is described

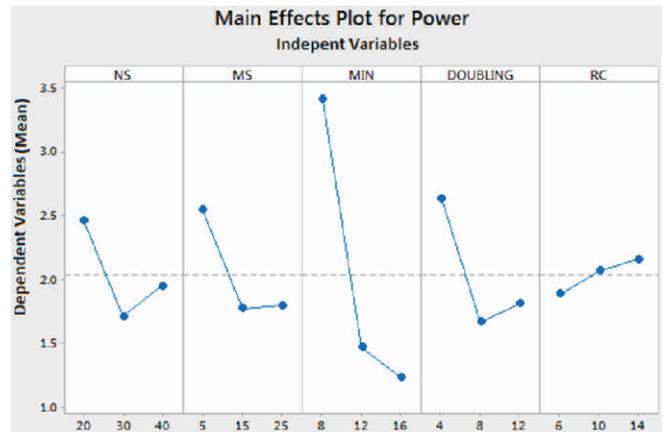

Figure 2. Main Effects Plot for SN ratio for Power

as follows:

- *Network Size:* The power consumption for an IoT network is high in case of 20 nodes and there is a drastic decrease in power consumption at a network size of 30 and a little increase in the power as the network size increases to 40 nodes.

- *Power Consumption:* The decrease in power consumption as the number of nodes increases in a network is due to the decrease in the number of DIO messages exchanges between nodes. As the number of nodes increases within a given area, it becomes easy to form a DODAG with less number of DIO exchanges. This happens up to a certain limit thereafter there will be congestion in the network which leads to the number of DIO messages to collide with each other. This results in retransmission of more number of DIO messages, which in turn results in an increase in power consumption.

- *Mobility Speed:* The mobility speed has the same effect as that of the network size. MIN value is inversely proportional to the power consumption (i.e, less MIN value higher the power consumption and more mobility speed lesser the power consumption).

The higher power consumption at fewer mobility speeds is due to the exchange of more DIO messages for forming the DODAG. As the mobility speed increases the DIO message exchanges will be reduced and hence the lesser power consumption.

- *DIO_MIN_INTERVAL:* Lower the min-interval values more will be the power consumption. As the min values are

increased there is a drastic decrease in the power consumption (i.e, at MIN=12). Further increase in the MIN value decreases the power consumption.

- *DIO_DOUBLING:* When doubling interval values increases the effect on power consumption decreases. In a similar way when doubling values decreases power consumption will be increased.

- *Redundancy Constant:* This is something difficult from the after-input parameter so far decreased. In this research article, it has a direct relationship with power consumption. From the above main effects plot, one can observe that Network Size, Mobility Speed, and DIO_MIN, DIO_DOUBLING are having an uneven relationship with power consumption (i.e, power consumption increases with a decrease in this parameter values and vice versa), where redundancy constant is having a direct relation with power consumption (i.e., power consumption increases with an increase in redundancy constant).

Figure 3 depicts the interaction plot of power which is a graphical tool to understand the interactions. In simple terms, it depicts the effect of an independent variable on the level of another independent variable. The analysis of Figure 3 is as follows:

The independent variables are said to have interaction if they are not parallel and do not have interaction if they are parallel to each other. For example, consider Row 1 and Column 5 in Figure 2: it gives the interaction of the independent variables (RC) on different levels (20, 30, and 40 nodes) of another independent variable NS. It is observed that the plots are almost parallel to each other and hence RC does not have any interaction with NS. Same is the case with RC and the other independent variables (MS, MIN, DOUBLING).

As the plots are almost parallel, there are no interactions with RC and MS, MIN, DOUBLING. As another example consider Row 1 and Column 3 (Interaction between levels of NS and various levels of MIN). It is clear from the graph that at low values of MIN, the power consumption is more and at higher values of MIN, the power consumption is less. From the same graph, observe that the plots are almost parallel for the values 30 and 40 of NS. Hence there are no interactions between them.

Continuing with the discussion consider Row 3 and Column 2 (interaction of different levels of MIN and various levels of MS). The plots are parallel and hence no interactions. Table 4 gives the results for various design points (simulation scenarios) for L27 OA.

Table 5 depicts the ANOVA table/Welch's test table where:

- *DF:* The amount of interaction present in the data can be treated as total Degrees of Freedom (DF). It shows how much information that a particular factor is used.

- *Seq SS:* Sequential sums of squares are measures of variation for different components of the model.

- *Ads MS:* Adjusted mean square measures how much variation a term explains regardless of the order they were entered.

- *F-Values:* The F-Value is the test statistic used to determine whether the term is associated with the response. A sufficiently large F-value indicates that the term is more significant.

- *P-Value:* It is the probability that measures the evidence against the NULL hypothesis. Lower probability provides stronger evidence against the NULL hypothesis. P-Value with the significance level (α=0.05) is used to find the significant factor.

*Case i:* If P-Value ≤ α: The difference between the sum of the means are statistically significant.

*Case ii:* If P-Values > α: The differences between the means are not statistically significant.

From Table 6 one can observe that factor C (MIN) is having highest F - Value (198.96) when compared with other factors and also P-Value < 0.05. Hence MIN is considered as the significant factor in power consumption of RPL. Considering the values of F - Value (2.64) and P-Value (0.102) > 0.05, Factor E (RC) is not considered to be a significant factor.

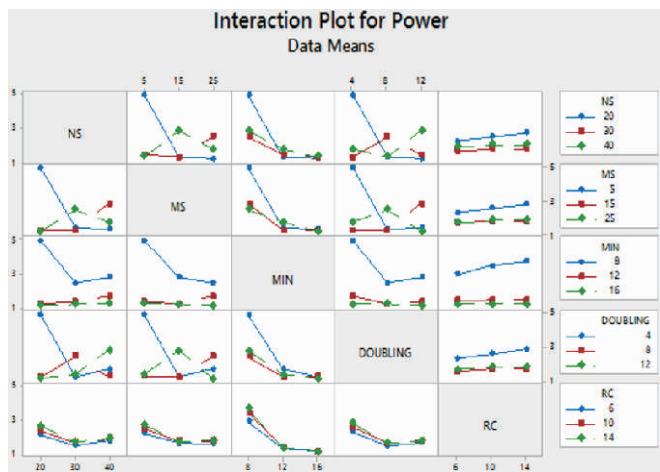

Figure 3. Interaction Plot for Power

| Design Point | Level of Factors | | | | | Power |
|---|---|---|---|---|---|---|
| | A | B | C | D | E | |
| 1 | 20 | 5 | 8 | 4 | 6 | 4.17 |
| 2 | 20 | 5 | 8 | 4 | 10 | 4.929 |
| 3 | 20 | 5 | 8 | 4 | 14 | 5.69 |
| 4 | 20 | 15 | 12 | 8 | 6 | 1.243 |
| 5 | 20 | 15 | 12 | 8 | 10 | 1.254 |
| 6 | 20 | 15 | 12 | 8 | 14 | 1.264 |
| 7 | 20 | 25 | 16 | 12 | 6 | 1.187 |
| 8 | 20 | 25 | 16 | 12 | 10 | 1.187 |
| 9 | 20 | 25 | 16 | 12 | 14 | 1.187 |
| 10 | 30 | 5 | 12 | 12 | 6 | 1.407 |
| 11 | 30 | 5 | 12 | 12 | 10 | 1.424 |
| 12 | 30 | 5 | 12 | 12 | 14 | 1.433 |
| 13 | 30 | 15 | 16 | 4 | 6 | 1.243 |
| 14 | 30 | 15 | 16 | 4 | 10 | 1.237 |
| 15 | 30 | 15 | 16 | 4 | 14 | 1.237 |
| 16 | 30 | 25 | 8 | 8 | 6 | 2.16 |
| 17 | 30 | 25 | 8 | 8 | 10 | 2.59 |
| 18 | 30 | 25 | 8 | 8 | 14 | 2.663 |
| 19 | 40 | 5 | 16 | 8 | 6 | 1.285 |
| 20 | 40 | 5 | 16 | 8 | 10 | 1.285 |
| 21 | 40 | 5 | 16 | 8 | 14 | 1.285 |
| 22 | 40 | 15 | 8 | 12 | 6 | 2.625 |
| 23 | 40 | 15 | 8 | 12 | 10 | 2.944 |
| 24 | 40 | 15 | 8 | 12 | 14 | 2.905 |
| 25 | 40 | 25 | 12 | 4 | 6 | 1.682 |
| 26 | 40 | 25 | 12 | 4 | 10 | 1.742 |
| 27 | 40 | 25 | 12 | 4 | 14 | 1.771 |

Table 4. Orthogonal Array with Experimental Results

| Factor | DF | Seq SS | Ads MS | F-Value | P-Value |
|---|---|---|---|---|---|
| A | 2 | 2.6184 | 1.3092 | 20.36 | 0 |
| B | 2 | 3.4758 | 1.7379 | 27.02 | 0 |
| C | 2 | 25.5925 | 12.7962 | 198.96 | 0 |
| D | 2 | 4.8743 | 2.4371 | 37.89 | 0 |
| E | 2 | 0.3392 | 0.1696 | 2.64 | 0.102 |
| Error | 16 | 1.0290 | 0.0643 | | |
| Total | 26 | 37.9292 | | | |

Table 5. ANOVA Table/Welch's Test Table

Table 6 gives the response table for SNR. The response table helps to identify the factors that have a significant effect on the output parameter. The factor with the largest delta values will be given rank 1, the factor with the second largest delta value is given rank 2, and so on. Hence from the table, the rows of MIN, DOUBLING, MS, NS, and RC are having ranges as 1, 2, 3, 4, and 5, respectively thus signifying the order of significant factor with respect to power.

| Level | NS | MS | MIN | DOUBLING | RC |
|---|---|---|---|---|---|
| 1 | -5.746 | -6.339 | -10.205 | -6.085 | -4.711 |
| 2 | -4.245 | -4.278 | -3.261 | -3.987 | -5.208 |
| 3 | -5.318 | -4.692 | -1.843 | -4.517 | -5.390 |
| Delta | 1.501 | 2.062 | 8.362 | 2.818 | 0.679 |
| Rank | 4 | 3 | 1 | 2 | 5 |

Table 6. Response Table for Signal to Noise Ratios

## Conclusion

In this research article, performance evaluation is done on RPL with five chosen parameters by using Taguchi Design of Experiments.

Cooja simulator is used for the evaluation in this research article. From the results discussed above, the power consumption in RPL has different significant factors, such as Network Size, Mobility Speed, DIO-MIN-INTERVAL, DIO-DOUBLING, and Redundancy Constant out of which DIO-MIN is considered to be the significant factor (i.e., DIO-MIN value has a significant effect on power consumption in RPL).

*Future Work*

One can focus on the application of optimization techniques such as Soft Computing/Hard Computing techniques on the significant factors identified in this research article and power consumption can be reduced thus paving a way to green technologies. In fact, an automated tool based on this can be developed, which means necessary changes in the value of DIO-MIN are automatically based on the input values, such as NS, MS, etc., for optimizing the RPL protocol.

## References

[1]. Ashton, K. (2009). That 'internet of things' thing. *RFID Journal*, 22(7), 97-114.

[2]. Box, G. E., Hunter, W. G., & Hunter, J. S. (2005). Statistics for Experimenters, 2nd Edition. *Wiley, New York.*

[3]. Coladon, T., Vučinić, M., & Tourancheau, B. (2015, August). Multiple redundancy constants with trickle. In *Personal, Indoor, and Mobile Radio Communications*

(PIMRC), 2015 IEEE 26th Annual International Symposium on (pp. 1951-1956). IEEE.

[4]. Dehnad, D. (1989). *Quality Control, Robust Design, and the Taguchi Method*. Wadsworth & Brooks/Cole, Pacific Grove, California.

[5]. Dunkels, A., Gronvall, B., & Voigt, T. (2004, November). Contiki- A lightweight and flexible operating system for tiny networked sensors. In *Local Computer Networks, 2004. 29th Annual IEEE International Conference on* (pp. 455-462). IEEE.

[6]. Ephrem, E. (2015). *The architecture of Wireless Sensor Networks*. Retrieved from http://servforu.blogspot.com.tr/2012/12/architecture-of-wireless sensor-networks.html

[7]. Evans, D. (2011). The Internet of Things: How the next evolution of the internet is changing everything. *CISCO White Paper*, 1(2011), 1-11.

[8]. Hazrat, A. (2012). *A performance evaluation of RPL in Contiki* (Masters Thesis, School of Computing, Blekinge Institute of Technology).

[9]. Kumar, S., Singh, D., & Chawla, M. (2011). Performance comparison of routing protocols in MANET varying network size. *International Journal of Smart Sensors and Ad Hoc Networks (IJSSAN)*, 1(2), 51-54.

[10]. Law, A. M., & Kelton, W. D. (2000). Simulation Modeling and Analysis. 3rd Edition. *Mc Graw Hill, New York*.

[11]. Mohamed, H., Lee, M. H., Sanugi, B., & Sarahintu, M. (2010). Taguchi approach for performance evaluation of routing protocols in mobile ad hoc networks. *Journal of Statistical Modeling and Analytics*, 1(2), 10-18.

[12]. Mohamed, H., Lee, M. H., Sarahintu, M., Salleh, S., & Sanugi, B. (2008). The use of Taguchi method to determine factors affecting the performance of destination sequence distance vector routing protocol in mobile ad hoc networks. *Journal of Mathematics and Statistics*, 4(4), 194-198.

[13]. Osterlind, F., Dunkels, A., Eriksson, J., Finne, N., & Voigt, T. (2006, November). Cross-level sensor network simulation with Cooja. In *Local Computer Networks, Proceedings 2006 31st IEEE Conference on* (pp. 641-648). IEEE.

[14]. Roy, R. K. (2001). *Design of Experiments using the Taguchi Approach: 16 Steps to Product and Process Improvement*. John Wiley & Sons.

[15]. Sahithi, K., & Setty, S.P. (2018). Taguchi Design of experiments for optimizing the performance of AODV routing protocol in MANETS. *International Journal of Engineering Research & Technology (IJERT)*, 7(4), 519-526.

[16]. Taguchi, G. (1986). *Introduction to quality engineering: designing quality into products and processes* (No. 658.562 T3).

[17]. Tsvetkov, T., & Klein, A. (2011). RPL: IPv6 routing protocol for low power and lossy networks. *Network*, 59.

[18]. Voice, A., Wilkins, A., Parambi, R., Oraiqat, I. (2007). *Factor Analysis and ANOVA*.

[19]. Winter, T., Thubert, P., Brandt, A., Hui, J., Kelsey, R., Levis, P., & Alexander, R. (2012). *RPL: IPv6 routing protocol for low-power and lossy networks* (No. RFC 6550).

## ABOUT THE AUTHORS

*Chandra Sekhar Sanaboina is presently pursuing Ph.D. under the guidance of Prof. Pallamsetty Sanaboina from Andhra University. He obtained his B.Tech in the Department of Electronics and Computer Science Engineering in 2005 and M.Tech in Computer Science and Engineering from Vellore Institute of Technology, Vellore, Tamil Nadu, India, in 2008. He has over 10 years of teaching experience and is currently working as an Assistant Professor in the Department of Computer Science and Engineering at JNTUK, Kakinada. His areas of interests include Wireless Sensor Networks, Internet of Things, Machine Learning, and Artificial Intelligence.*

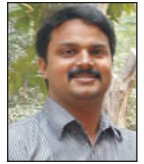

*Dr. Pallamsetty Sanaboina is currently working as a Professor in the Department of Computer Science and Systems Engineering of Andhra University, Visakhapatnam. He has completed his Ph.D. in the area of Wireless Sensor Networks from Andhra University, Visakhapatnam, Andhra Pradesh, India. He has twenty-eight years of Teaching and Research Experience. He has over 200 publications in International Journals and Conferences of repute. More than 200 Master's Students and 18 Doctoral Students had completed their Degree so far under his supervision, and currently supervising 8 Doctoral candidates in the areas of Mobile Ad-hoc Networks, Wireless Sensor Networks, Ubiquitous Computing, and Web of Things, which are also his research interests.*

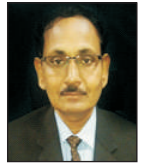